# Plasmonic nature of periodic surface structures following single laser pulse irradiation

P.N. Terekhin[1,*,#], J. Oltmanns[2,#], A. Blumenstein[2], D.S. Ivanov[3], F. Kleinwort[2], M.E. Garcia[4], B. Rethfeld[1], J. Ihlemann[2], and P. Simon[2]

[1]*Department of Physics and OPTIMAS Research Center, Technische Universität Kaiserslautern, Erwin-Schrödinger-Strasse 46, 67663 Kaiserslautern, Germany*

[2]*Institut für Nanophotonik Göttingen e.V., Hans-Adolf-Krebs-Weg 1, 37077 Göttingen, Germany*

[3]*Quantum Electronics Division, Lebedev Physical Institute, 119991 Moscow, Russia*

[4]*Theoretical Physics Department, University of Kassel, 34132 Kassel, Germany*

[#]*These authors contributed equally to this work*

**Abstract**

Understanding the mechanisms and controlling the possibilities of surface nanostructuring is of crucial interest from fundamental and practical perspectives. Here we report a direct experimental observation of laser-induced periodic surface structures (LIPSS) formed near a predesigned gold step edge following single-pulse femtosecond laser irradiation. A hybrid atomistic-continuum model fully supports experimental observations. We identify two key components of single-pulse LIPSS formation: excitation of surface plasmon polaritons and material reorganization. Our results lay the foundation towards simple and efficient single laser pulse micromachining.

*Corresponding author.

E-mail address: terekhin@physik.uni-kl.de (P.N. Terekhin)

Controlling radiation-induced fabrication of surface nanostructures is highly demanded for the development of advanced nanoscale devices. Coupling femtosecond laser pulses to surface nanoreliefs can lead to the spontaneous formation of laser-induced periodic surface structures (LIPSS), which have gained a broad spectrum of potential applications in industry, medicine, biology, and optics [1-3] due to their unique properties. LIPSS are usually formed in a multi-pulse regime either at normal or at oblique incidence [3-7]. The surface morphology is continuously changing from pulse to pulse. This has a crucial influence on the energy absorption of subsequent laser irradiation. Before tackling an intricate question about LIPSS formation in a multi-pulse regime, it is necessary first to reveal the mechanisms for the LIPSS formation due to a single laser pulse for efficient and controllable surface engineering.

Although there is an overwhelming number of publications on multi-pulse LIPSS, only few studies exist on LIPSS formation with a single-pulse irradiation [8-14]. Among a number of theories proposed to explain the formation of LIPSS [3,7,13,15-21], excitation of surface plasmon polaritons (SPP) and their interference with the incoming light wave is regarded to be the most promising hypothesis for the physical origin of the LIPSS morphology. SPP are extensively used in plasmonics at low pulse energy for investigation of the extraordinary light transmission [22], dielectric-loaded SPP waveguides [23], quantum plasmonic interference effects [24], on-chip photonic devices [25] and nano-antennas [26,27]. However, little attention has been given up to date to verify SPP influence on LIPSS formation at high laser fluences.

LIPSS formation on material surfaces was investigated in different works [13,17] in the framework of the two-temperature model (TTM) [28,29], where it was assumed that electrons absorb the laser energy in accordance with the source term profile. In addition, a combined electromagnetic and thermohydrodynamic approach [21] was developed. However, under strong superheating and the presence of strong pressure gradients, the developing new phase nuclei can be as small as a few times the interatomic distance. Therefore, the crystal structure must be considered. Hence, continuum models are not applicable to any situation, whereas molecular dynamics (MD)-based simulations are of more general validity. So far, arbitrary periodic functions in the lateral *x*-direction were assumed for the source term to imitate the laser energy absorption in the frames of the TTM or MD-TTM models [13,17,19,20]. As a result, such models allow to reveal relaxation dynamics of laser-induced periodic energy deposition, but they cannot inherently predict the SPP periodicity and decay length, which will influence the final morphology of LIPSS.

In this Letter, we present a clear demonstration of the SPP nature of LIPSS formation following single laser pulse irradiation of a step edge structure on a gold sample. In addition, our MD-TTM simulations, including SPP excitation and subsequent interference of the SPP field with the laser field [30], provide a complete description of the experimentally obtained structures.

SPP waves can be induced at the interface between a dielectric (air) and a conductor and propagate along it. For the excitation of the SPP, two conditions have to be met. First, the geometrical structure of the irradiated sample should provide the missing wave vector between the photon (laser pulse) and the SPP to fulfill the conservation of linear momentum. Therefore, it is possible to launch SPP waves on a sample with a step edge [31,32] or rough surfaces [33]. The SPP coupling can also be realized with the help of a prism or grating [33]. Second, the dielectric function of the conducting medium, at the interface of which the SPP can be excited, must satisfy the condition on plasmonic activity that is $Re(\varepsilon) < 0$, where $\varepsilon$ is the dielectric function [33]. Simultaneous fulfillment of both conditions for the SPP excitation can be guaranteed by a step edge structure used in our study. The schematic irradiation geometry is shown in Fig. 1(a).

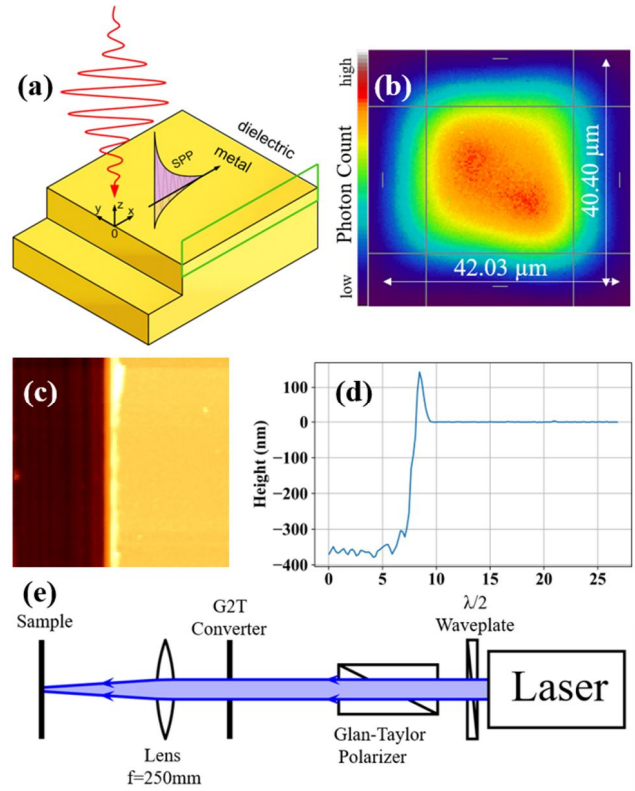

**Fig. 1.** (a) The schematic irradiation geometry with the SPP excitation. The green frame shows the simulation region for the MD-TTM. (b) The beam profile in the surface plane. (c) AFM recording of a trench on the gold sample and (d) the average height profile measured by the AFM. (e) The experimental setup, containing the laser source, a λ/2 wave plate and a Glan-Taylor polarizer to control the polarization of the beam and a Gaussian to top-hat beam shaper (G2T) with a focusing lens (f = 250 mm) to produce a homogeneous, square-shaped laser spot (~ 40 µm at 1/e² level, see figure part (b)) on the sample.

For this given structure, one can univocally determine the SPP excitation and its position [31]. Generally, the SPP will be excited at both sides of the step edge, but here we are interested in the structures on the top of Au with positive *x*-coordinate values. An analytical function for the source term (the total input power per volume) $Q_{total}(\mathbf{r},t)$ with $\mathbf{r}$ given by (*x,y,z*) in the frames of the TTM approach has been derived by explicit calculation of the electromagnetic fields of the laser and the SPP and their interference [30]. It turns out to consist of a sum of different interference contributions (for details, see in the Supplemental Material [34]). The dielectric function was assumed to be constant and taken from an experiment [35] for the evaporated gold sample.

We apply a hybrid atomic-continuum model (MD-TTM) [29,36,37] to describe LIPSS formation after single-pulse laser irradiation. In this model, the molecular dynamics approach is used to describe the transient states of matter due to the laser-induced structural dynamics with atomic precision, whereas the TTM part of the model accounts for the laser energy absorption, fast electron heat conduction, and strong laser-induced electron-phonon non-equilibrium in the continuum state. In contrast to previous studies [13,17,19,20], the description of the laser energy absorption is given by the recently developed source term $Q_{\text{total}}(\mathbf{r},t)$ that includes the effect of SPP excitation on the laser energy redistribution across the surface [30]. We verify our findings by means of the direct comparison of the simulation results with the experimental data. The detailed description of the MD-TTM model, computational cells and boundary conditions are given elsewhere [37], whereas an advanced model on the electron conductivity [38] was implemented for massive parallel simulations of the laser-deposited energy dissipation in the irradiated target.

In our experiments, the sample surface consists of a gold layer on a glass substrate with a thickness of about 350 nm [Fig. 1(d)]. The root mean square (rms) surface roughness on the top of the gold sample was measured to be $R_{\text{rms}}$=1 nm. Such a small roughness ensures that SPP are excited only at the position of the step edge. The polarization of the laser beam is typically set to be perpendicular to the step edge (p-polarization). In addition, we have performed experiments with the laser polarization being parallel to the step edge (s-polarization), where LIPSS are not observed. Therefore, this ensures that the excitation of any other surface waves, for instance, quasi-cylindrical waves, has a negligible contribution to single-pulse LIPSS formation for a given sample geometry. Moreover, it was shown that quasi-cylindrical waves are mainly important for sub-wavelength indentations [39,40].

As the experimental representation of the step edge, a trench was mechanically created by scratching the surface before irradiation with the help of a sharp razor blade by loading it onto the surface using a spring to maintain a constant force. The edges of this trench are the equivalent of the step edge. The sample was moved with a constant velocity via a motorized linear stage to provide a steady trench. The resulting profile can be seen in Fig. 1 (c,d). The uplifting walls at the sides of the trench have a height of 100-200 nm relative to the gold surface. The laser beam was centered at the irradiated trench edge.

All experimental results were produced using the third harmonic radiation of a Light Conversion Pharos PH-1-20W at 343 nm wavelength with a spectral width of about 1 nm. At the fundamental wavelength of 1030 nm, the pulse duration is 250 fs. A sketch and a further description of the experimental setup are seen in Fig. 1(e). The beam profile in the sample plane is displayed in Fig. 1(b). All experiments were done using one single pulse per irradiation at normal incidence. A series of single pulses were given on different positions along the trench, while the fluences were varied from spot to spot. The sample was evaluated by a scanning electron microscope (SEM) and an atomic force microscope (AFM).

The threshold of visible periodic patterns with a period of ~ 343 nm appears on a gold surface after laser irradiation of the step edge with a fluence of 160 mJ/cm². Increasing the fluence enhances the visibility of the structures. Prominent LIPSS at fluences 172 mJ/cm² and 192 mJ/cm² are shown in Fig. 2. In the investigated fluence range (130-350 mJ/cm$^2$), we observe LIPSS at fluences 160-270 mJ/cm², where the LIPSS periodicity stays constant. Above 270 mJ/cm², we do not detect structures on a gold surface in the final morphology. A possible explanation is that they disappear upon ablation due to removing the SPP-affected depth inside the sample.

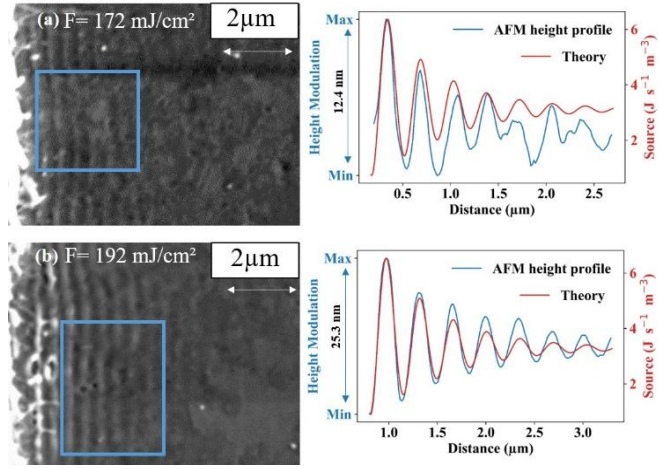

**Fig. 2.** SEM picture of LIPSS structures at the edge of the trench irradiated with the fluences (a) F=(172± 9) mJ/cm² and (b) F=(192±10) mJ/cm². The height profile of LIPSS is measured by AFM averaged over the sections indicated in the SEM pictures. The theoretical curves of the source function are also included in the height measurement.

In the single-pulse experiments, we observe low-spatial-frequency LIPSS (LSFL), which are oriented orthogonal to the beam polarization and have a periodicity close to the laser wavelength. We show in Fig. 2 that material removal is not necessarily required for a single-pulse LSFL formation. To the best of our knowledge, we have manufactured LSFL with the smallest periodicity reported so far for the single-pulse irradiation of metals in the air. The only exception we are aware of is LSFL structures on Cu within the same range of periodicity (300 ± 40) nm [12].

To simulate the LIPSS formation, we applied MD-TTM with a single 250 fs laser pulse and an incident fluence of 130 mJ/cm². The sample size was 4000 nm × 10 nm × 200 nm in $x$, $y$, and $z$ directions, respectively, consisting of 375 million atoms. The position of the step edge is located in our simulations at the center, where $x = 0$ nm, and periodic boundary conditions are set in the x-direction at ± 2000 nm, a distance large enough from the center to neglect the modulations in laser energy deposition due to the SPP. The Non-Reflective Boundary (NRB) conditions [41] were applied at a depth of 150 nm, where no phase transformations are expected, but the laser-induced pressure waves must be absorbed. While imitating the material below in such a way and avoiding unnecessary MD integrations, an ordinary TTM model was solved beneath NRB for both electrons and phonons. In the simulation, we assume the SPP coupling parameter $\beta = 0.30$ and the phase $\delta = 238.68°$ [30] in the implemented laser source (see in the Supplemental Material [34]). These parameters were chosen based on preliminary calculations on the laser energy deposition coupled with the energy conservation law. The electronic temperature reaches its maximum value of 14500 K and performs roughly five modulations before it decays (the first two oscillations are visible in Fig. S1(a) in the Supplemental Material [34]). Followed by the laser energy absorption, the electron-phonon equilibration takes place within the next 10 ps. The maximum lattice temperature of 3500 K is achieved in the proximity of the surface near the step edge (see Fig. S1(b) in the Supplemental Material [34]).

    The simulation results are shown in Fig. 3 as a sequence of the atomic snapshots of the target evolution. The material reorganization (surface uplift) of LIPSS formation occurs at 120 ps (see Fig. 3) when the nucleation of internal voids, due to the relaxation of the laser-induced stresses, indicates the onset of spallation. While structures are forming, the resolidification process takes place. This initially can be seen in Fig. 3 at 300 ps, 500 ps, and 1000 ps as a propagation of the solid-liquid interface towards the free surface due to the heterogeneous mechanism of solidification. Later, at 1000 ps, the resolidification process is complemented with nucleation of solid phase inside the liquid due to homogeneous solidification mechanism, when the strong cooling rate results in the local temperature drop significantly below the melting point. The structure closest to the step edge grows up to 70 nm by the time of 1000 ps, while the third from the step edge structure has a height of 25 nm (see Fig. 3). This is in good agreement with the experimental height profile shown in Fig. 2(b). The AFM measurement does not capture the first two periods, and the observed height profile starts with a modulation depth of about 25 nm. The new surface level is elevated above its initial position before the

irradiation by 2-3 nm due to material thermal expansion and formation of a number of dislocation planes, visible as light blue lines in Fig. 3 beginning from 300 ps (see Fig. S2 in the Supplemental Material [34] with the zoomed atomic snapshot at 1000 ps).

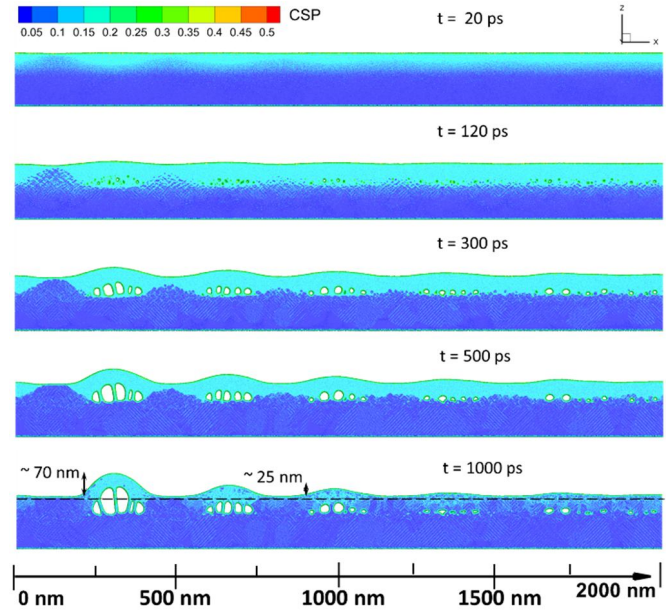

**Fig. 3.** Atomic snapshots of the Au target for the case of an incident fluence of 130 mJ/cm$^2$ are depicted for the times of 20 ps, 120 ps, 300 ps, 500 ps and 1000 ps after the pulse. The atoms are colored by the Central Symmetry Parameter (CSP) for distinguishing the local crystal structure as follow: crystal < 0.08 < defects (dislocations) < 0.11 < liquid < 0.25 < surfaces < 0.50 < vapor (free atoms). The dashed line at the last snapshot indicates the sample surface before the irradiation.

From the analysis of the simulation results, we can state that the mechanism of the formation of periodic structures after the laser pulse deposition is in general similar to that discussed in Ref. [37]. This means that due to the periodic lattice heating, the periodic structures (LIPSS) formation occurs because of the nucleation of voids in the bulk of the material with their periodic concentration corresponding to the peaks of the SPP-assisted laser energy distribution. Owing to the laser heating process under the conditions of internal stress confinement [29], the subsequent relaxation of the laser-induced internal stresses generates dislocation planes along with the propagation of a strong pressure wave (~15 GPa, see Fig. S3 in the Supplemental Material [34]) through the molten material. The establishment of a hydrodynamic motion is responsible for the elevation of the forming structure [42]. The solidification process of the growing LIPSS is provided by the fast electron heat conduction towards the bulk material in *z*-direction [42].

We find excellent agreement between experiments and simulations for LIPSS periodicity: $\lambda_{\text{LIPSS}}^{\text{exp}} = (338 \pm 10)\,\text{nm}$ and $\lambda_{\text{SPP}}^{\text{theory}} = 344\,\text{nm}$. The spatial decay of the periodic structures away from the step edge is also in accordance with the decay of the modulated absorbed laser energy in the source term (see Fig. 2). However, the experimental structures decay a bit slower in space. The reason could be the non-constant dielectric function [43], resulting in an increased SPP propagation length and, therefore, its decay length. In addition, the limited size of the computational cell and utilization of NRB boundaries designed for the absorption of planar pressure waves could also account for an underestimated threshold of the growing LIPSS structures. More accurate Langevin-NRB, aimed on the absorption of non-planar laser-induced pressure waves due to periodically heated material surface, have been recently suggested in Ref. [19] and will be used in our following research. Thus, although the data calculated for 130 mJ/cm² match the experimental results for 192 mJ/cm², the direct comparison of our modeling results with our experimental findings provides a full understanding of the mechanism of the structure formation in a semi-quantitative manner. However, due to the reasons mentioned above, additional investigations are necessary for a precise determination of the SPP coupling parameter β and hence a fully quantitative description of the LIPSS formation.

Based on our findings, we can elucidate the key steps leading to LIPSS formation after single-shot femtosecond irradiation of a predesigned step edge feature on gold. First, SPP waves are responsible for the periodic laser energy deposition and seed an opportunity that LIPSS will be formed. Although SPP can be excited at all fluences, LIPSS can be observed only in a specific laser fluence range. It means that the laser energy must be high enough to create visible structures, but it has to be below the fluence value, at which LIPSS will be destroyed due to ablation (material removal process). The ablation threshold will depend on the laser wavelength and the material's properties. Second, the final LIPSS relief is governed by the material elevation due to relaxation of the laser-induced stresses, triggering the spallation mechanism (mechanical rupture of the material with the formation of internal voids). Whereas, the solidification process is assisted with fast cooling due to the electron thermal conduction process. Thus, the evolution of the periodically heated material will define the LIPSS morphology.

In summary, in this Letter we demonstrate the SPP nature of LIPSS through irradiation of a step edge feature on Au due to a single laser pulse. We have identified two key components of LIPSS formation: SPP excitation, responsible for laser-deposited energy modulation, and material reorganization, responsible for the final LIPSS morphology. The comprehensive MD-TTM simulations with the laser-SPP source term combined with experimental measurements enabled us to describe LIPSS formation and their evolution. We have found an excellent agreement in the periodicity and the decay of LIPSS between experimental results and simulations after irradiation of a step edge on gold at 343 nm. Future experiments with manipulating the laser fluence and the beam shape can pave the way for LIPSS design with demanded height profiles and periodicity. A powerful MD-TTM tool with a previously developed analytical source term is able to describe the underlying physical processes on an atomic scale with further potential for applications in nanostructuring and plasmonics.

**Acknowledgments**

We acknowledge the financial support of the Deutsche Forschungsgemeinschaft projects RE1141/14-2, IH 17/18-2, GA465/15-2. We are grateful to S.T. Weber, O. Benhayoun, K.P. Migdal and N.A. Inogamov for fruitful discussions. The MD-TTM calculations were performed at Lichtenberg Super Computer Facility TU-Darmstadt (Germany). Some simulations were executed on the high-performance cluster "Elwetritsch" through the projects TopNano and Mulan at the TU Kaiserslautern, which is a part of the "Alliance of High Performance Computing Rheinland-Pfalz". P.N.T. and B.R. kindly acknowledge the support of Regionales Hochschulrechenzentrum Kaiserslautern.

**References**

1. J. Bonse, S. Höhm, S.V. Kirner, A. Rosenfeld, and J. Krüger, Laser-induced Periodic Surface Structures (LIPSS) – A Scientific Evergreen, IEEE J. Sel. Top. Quantum Electron. **23**, 9000615 (2017).

2. C. Florian, S. V. Kirner, J. Krüger, and J. Bonse, Surface functionalization by laser-induced periodic surface structures, J. Laser Appl. **32**, 022063 (2020).

3. J. Bonse and S. Gräf, Maxwell Meets Marangoni - A Review of Theories on Laser-Induced Periodic Surface Structures, Laser Photonics Rev. **14**, 2000215 (2020).

4. A.Y. Vorobyev, V. S. Makin, and Chunlei Guo, Brighter Light Sources from Black Metal: Significant Increase in Emission Efficiency of Incandescent Light Sources, Phys. Rev. Lett. **102**, 234301 (2009).

5. Y. Liu, Y. Brelet, Z. He, L. Yu, S. Mitryukovskiy, A. Houard, B. Forestier, A. Couairon, and A. Mysyrowicz, Ciliary White Light: Optical Aspect of Ultrashort Laser Ablation on Transparent Dielectrics, Phys. Rev. Lett. **110**, 097601 (2013).

6. C.A. Zuhlke, G.D. Tsibidis, T. Anderson, E. Stratakis, G. Gogos, and D.R. Alexander, Investigation of femtosecond laser induced ripple formation on copper for varying incident angle, AIP Adv. **8**, 015212 (2018).

7. H. Zhang, J.P. Colombier, and S. Witte, Laser-induced periodic surface structures: Arbitrary angles of incidence and polarization states, Phys. Rev. B **101**, 245430 (2020).

8. R.D. Murphy, B. Torralva, D. P. Adams, and S. M. Yalisove, Laser-induced periodic surface structure formation resulting from single-pulse ultrafast irradiation of Au microstructures on a Si substrate, Appl. Phys. Lett. **102**, 211101 (2013).

9. R.D. Murphy, B. Torralva, D. P. Adams, and S. M. Yalisove, Polarization dependent formation of femtosecond laser-induced periodic surface structures near stepped features, Appl. Phys. Lett. **104**, 231117 (2014).

10. E.L. Gurevich, On the influence of surface plasmon-polariton waves on pattern formation upon laser ablation, Appl. Surf. Sci. **278**, 52 (2013).

11. E.L. Gurevich, S.V. Gurevich, Laser Induced Periodic Surface Structures induced by surface plasmons coupled via roughness, Appl. Surf. Sci. **302**, 118 (2014).

12. S. Maragkaki, T.J.-Y. Derrien, Y. Levy, N.M. Bulgakova, A. Ostendorf, E. L. Gurevich, Wavelength dependence of picosecond laser-induced periodic surface structures on copper, Appl. Surf. Sci. **417**, 88 (2017).

13. E. L. Gurevich, Y. Levy, and N. M. Bulgakova, Three-Step Description of Single-Pulse Formation of Laser-Induced Periodic Surface Structures on Metals, J. Nanomater. **10**, 1836 (2020).

14. N. Ackerl, and K. Wegener, Observation of single ultrashort laser pulse generated periodic surface structures on linelike defects, J. Laser Appl. **32**, 022049 (2020).

15. J.E. Sipe, J.F. Young, J.S. Preston, and H.M. van Driel, Laser-induced periodic surface structure. I. Theory, Phys. Rev. B **27**, 1141 (1983).

16. J. Bonse, A. Rosenfeld, and J. Krüger, On the role of surface plasmon polaritons in the formation of laser-induced periodic surface structures upon irradiation of silicon by femtosecond-laser pulses, J. Appl. Phys. **106**, 104910 (2009).

17. Y. Levy, N.M. Bulgakova, and T. Mocek, Laser-induced periodic surface structures formation: investigation of the effect of nonlinear absorption of laser energy in different materials, Proc. SPIE **10228**, 102280T (2017).

18. O. Varlamova, J. Reif, S. Varlamov, and M. Bestehorn, The laser polarization as control parameter in the formation of laser-induced periodic surface structures: Comparison of numerical and experimental results, Appl. Surf. Sci. **257**, 5465 (2011).

19. M. V. Shugaev, I. Gnilitskyi, N. M. Bulgakova, and L.V. Zhigilei, Mechanism of single-pulse ablative generation of laser-induced periodic surface structures, Phys. Rev. B **96**, 205429 (2017).

20. C.-Y. Shih, I. Gnilitskyi, M. V. Shugaev, E. Skoulas, E. Stratakis and L. V. Zhigilei, Effect of a liquid environment on single-pulse generation of laser induced periodic surface structures and nanoparticles, Nanoscale **12**, 7674 (2020).

21. A. Rudenko, A. Abou-Saleh, F. Pigeon, C. Mauclair, F. Garrelie, R. Stoian, J.P. Colombier, High-frequency periodic patterns driven by non-radiative fields coupled with Marangoni convection instabilities on laser-excited metal surfaces, Acta Mater. **194**, 93 (2020).

22. L. Aigouy, P. Lalanne, J. P. Hugonin, G. Julié, V. Mathet, and M. Mortier, Near-Field Analysis of Surface Waves Launched at Nanoslit Apertures, Phys. Rev. Lett. **98**, 153902 (2007).

23. T. Leißner, C. Lemke, J. Fiutowski, J.W. Radke, A. Klick, L. Tavares, J. Kjelstrup-Hansen, H.-G. Rubahn, and M. Bauer, Morphological Tuning of the Plasmon Dispersion Relation in Dielectric-Loaded Nanofiber Waveguides, Phys. Rev. Lett. **111**, 046802 (2013).

24. A. Safari, R. Fickler, E. Giese, O.S. Magaña-Loaiza, R.W. Boyd, and I. De Leon, Measurement of the Photon-Plasmon Coupling Phase Shift, Phys. Rev. Lett. **122**, 133601 (2019).

25. V. Kravtsov, S. AlMutairi, R. Ulbricht, A. Ryan Kutayiah, A. Belyanin, and M.B. Raschke, Enhanced Third-Order Optical Nonlinearity Driven by Surface-Plasmon Field Gradients, Phys. Rev. Lett. **120**, 203903 (2018).

26. T. Zang, H. Zang, Z. Xi, J. Du, H. Wang, Y. Lu, and P. Wang, Asymmetric Excitation of Surface Plasmon Polaritons via Paired Slot Antennas for Angstrom Displacement Sensing, Phys. Rev. Lett. **124**, 243901 (2020).

27. V. Bruno, C. DeVault, S. Vezzoli, Z. Kudyshev, T. Huq, S. Mignuzzi, A. Jacassi, S. Saha, Y. D. Shah, S.A. Maier, D.R.S. Cumming, A. Boltasseva, M. Ferrera, M. Clerici, D. Faccio, R. Sapienza, and V.M. Shalaev, Negative Refraction in Time-Varying Strongly Coupled Plasmonic-Antenna–Epsilon-Near-Zero Systems, Phys. Rev. Lett. **124**, 043902 (2020).

28. S.I. Anisimov, B.L. Kapeliovich, and T.L. Perel'man, Electron Emission from Metal Surfaces Exposed to Ultrashort Laser Pulses, Zh. Eksp. Teor. Fiz. **66**, 776 (1974).

29. B. Rethfeld, D.S. Ivanov, M.E. Garcia and S.I. Anisimov, Modelling ultrafast laser ablation, J. Phys. D: Appl. Phys. **50**, 193001 (2017).

30. P.N. Terekhin, O. Benhayoun, S.T. Weber, D.S. Ivanov, M.E. Garcia, B. Rethfeld, Influence of surface plasmon polaritons on laser energy absorption and structuring of surfaces, Appl. Sur. Sci. **512**, 144420 (2020).

31. C. Lemke, T. Leißner, A. Klick, J. Fiutowski, J.W. Radke, M. Thomaschewski, J. Kjelstrup-Hansen, H.-G. Rubahn, M. Bauer, The complex dispersion relation of surface plasmon polaritons at gold/para-hexaphenylene interfaces, Appl. Phys. B **116**, 585 (2014).

32. A. Klick, S. de la Cruz, C. Lemke, M. Großmann, H. Beyer, J. Fiutowski, H.-G. Rubahn, E.R. Méndez, M. Bauer, Amplitude and phase of surface plasmon polaritons excited at a step edge, Appl. Phys. B **122**, 79 (2016).

33. S.A. Maier, Plasmonics, Fundamentals and Applications, Springer, Berlin, 2007.

34. See Supplemental Material at URL for additional theoretical analysis, which includes Ref. [30].

35. R.L. Olmon, B. Slovick, T.W. Johnson, D. Shelton, S.-H. Oh, G.D. Boreman, and M.B. Raschke, Optical dielectric function of gold, Phys. Rev. B **86**, 235147 (2012).

36. D.S. Ivanov and L.V. Zhigilei, Combined atomistic-continuum modeling of short-pulse laser melting and disintegration of metal films, Phys. Rev. B **68**, 064114 (2003).

37. D.S. Ivanov, V.P. Lipp, A. Blumenstein, F. Kleinwort, V.P. Veiko, E. Yakovlev, V. Roddatis, M.E. Garcia, B. Rethfeld, J. Ihlemann, and P. Simon, Experimental and Theoretical Investigation of Periodic Nanostructuring of Au with Ultrashort UV Laser Pulses near the Damage Threshold, Phys. Rev. Appl. **4**, 064006 (2015).

38. Yu.V. Petrov, N.A. Inogamov, S.I. Anisimov, K.P. Migdal, V.A. Khokhlov, and K.V. Khishchenko, Thermal conductivity of condensed gold in states with the strongly excited electron subsystem, J. Phys. Conf. Ser. **653**, 012087 (2015).

39. P. Lalanne and J.P. Hugonin, Interaction between optical nano-objects at metallo-dielectric interfaces, Nature Phys. **2**, 551 (2006).

40. P. Lalanne, J.P. Hugonin, H.T. Liu, B. Wang, A microscopic view of the electromagnetic properties of sub-λ metallic surfaces, Surf. Sci. Rep. **64**, 453 (2009).

41. C. Schäfer, H. M. Urbassek, L. V. Zhigilei, and B. J. Garrison, Pressure-transmitting boundary conditions for molecular-dynamics simulations, Comput. Mater. Sci. **24**, 421 (2002).

42. D.S. Ivanov, B.C. Rethfeld, G.M. O'Connor, T.J. Glynn, A.N. Volkov, and L.V. Zhigilei, The Mechanism of Nanobump Formation in Femtosecond Pulse Laser Nanostructuring of Thin Metal Films, Appl. Phys. A. **92**, 791 (2008).

43. A. Blumenstein, E. S. Zijlstra, D. S. Ivanov, S. T. Weber, T. Zier, F. Kleinwort, B. Rethfeld, J. Ihlemann, P. Simon, and M. E. Garcia, Transient optics of gold during laser irradiation: From first principles to experiment, Phys. Rev. B **101**, 165140 (2020).

# Supplemental Material:

# Plasmonic nature of periodic surface structures following single laser pulse irradiation

P.N. Terekhin[1,*,#], J. Oltmanns[2,#], A. Blumenstein[2], D.S. Ivanov[3], F. Kleinwort[2], M.E. Garcia[4], B. Rethfeld[1], J. Ihlemann[2], and P. Simon[2]

[1]Department of Physics and OPTIMAS Research Center, Technische Universität Kaiserslautern, Erwin-Schrödinger-Strasse 46, 67663 Kaiserslautern, Germany

[2]Institut für Nanophotonik Göttingen e.V., Hans-Adolf-Krebs-Weg 1, 37077 Göttingen, Germany

[3]Quantum Electronics Division, Lebedev Physical Institute, 119991 Moscow, Russia

[4]Theoretical Physics Department, University of Kassel, 34132 Kassel, Germany

*Corresponding author e-mail address: terekhin@physik.uni-kl.de (P.N. Terekhin)

[#]These authors contributed equally to this work.

**The theoretical model for the description of the laser energy deposition**

We describe the energy absorption upon irradiation of a gold sample by a single femtosecond laser pulse following Ref. [1]. An analytical function for the source term in the frames of the MD-TTM approach has been derived by explicit calculation of the interference of the SPP fields and the laser fields. However, in this work, we have used a flat-top incident beam shape instead of a Gaussian shape, applied in Ref. [1], to directly compare simulations and experiments. In this case, the expression for the source term has a form

$$Q_{\text{total}}(\mathbf{r},t,\beta,\delta) = Q_{\text{las-las}}(\mathbf{r},t) + Q_{\text{las-SPP}}(\mathbf{r},t,\beta,\delta) + Q_{\text{SPP-SPP}}(\mathbf{r},t,\beta,\delta), \quad \text{(S1)}$$

where the first term (the coordinate system is marked in Fig.1(a) of the main text)

$$Q_{\text{las-las}}(\mathbf{r},t) = F_{\text{inc}} \frac{8 k_0 n_m k_m}{(n_m+1)^2 + k_m^2} e^{2 k_0 k_m z} \Phi_1(x,y) \frac{1}{\tau}\sqrt{\frac{\sigma}{\pi}} e^{-\sigma \frac{(t-t_0)^2}{\tau^2}} \quad \text{(S2)}$$

describes the laser-laser interference with $F_{inc}$ being the incident peak laser fluence

$$F_{inc} = \frac{E_{pulse}}{S_{beam}}, \quad (S3)$$

where $E_{pulse}$ is the total pulse energy and $S_{beam} = a_{beam}^2$ is the beam area [see Fig. 1(b) in the main text] with $a_{beam}$ being the linear beam size at 1/e² level. $k_0 = \omega/c$ is the wave vector of light, $\omega$ is the laser angular frequency, $c$ is the speed of light, $n_m$ and $k_m$ are the real and imaginary parts of the complex refractive index of metal $\tilde{n}_m = n_m + ik_m$, $\tau$ is a pulse duration at FWHM (full width at half maximum), $t_0 = 2.5\tau$ is chosen as a location of the pulse maximum and $\sigma = 4\ln 2$. The function

$$\Phi_1(x,y) = \theta\left(-x + \frac{a_{beam}}{2}\right)\theta\left(x + \frac{a_{beam}}{2}\right)\theta\left(-y + \frac{a_{beam}}{2}\right)\theta\left(y + \frac{a_{beam}}{2}\right) \quad (S4)$$

is responsible for the flat-top laser shape.

The second term of the Eq. (S1) stands for the laser-SPP interference:

$$Q_{las\text{-}SPP}(\mathbf{r},t,\beta,\delta) = \beta F_{inc} \frac{2\left[f_1 \cos(f_3(x,z,\delta)) + f_2 \sin(f_3(x,z,\delta))\right]}{(n_m+1)^2 + k_m^2} \Phi_1(x,y) G_1(x,t) \\ \times \exp\left((k_0 k_m + k'_{z,m})z - k''_x x\right), \quad (S5)$$

where $\beta$ is the SPP coupling efficiency, which is equal to the ratio of moduli of the amplitude of the SPP magnetic field to the amplitude of the incident magnetic field of a laser at the position of the step edge ($x=0,y=0,z=0$); $\delta$ is the phase difference between the incident beam and the excited SPP, which describes the shift of the source term profile along the lateral distance [1]. $k'_{z,m}$ is the real part of the wave vector of the SPP in the metal in the z-direction $k_{z,m} = k'_{z,m} + ik''_{z,m}$; $k''_x$ is the imaginary part of the wave vector of the SPP in the x-direction $k_x = k'_x + ik''_x$ and the function $G_1(x,t)$ is defined as

$$G_1(x,t) = \frac{1}{\tau}\sqrt{\frac{\sigma}{\pi}} e^{-\sigma\frac{(t-t_0)^2}{2\tau^2}} e^{-\sigma\frac{(x-v_{g,SPP}(t-t_0))^2}{2v_{g,SPP}^2\tau^2}}, \quad (S6)$$

where $v_{g,SPP} = d\omega/dk'_x$ is the group velocity of the SPP. The values $f_1 - f_3$ including $f_4 - f_7$ are the following:

$$f_1 = \frac{-k''_x f_4 + k'_x f_5}{k_0 |\varepsilon_m|^2} - (k'_{z,m} + k_0 k_m) f_6 - (k''_{z,m} + k_0 n_m) f_7, \tag{S7}$$

$$f_2 = -\frac{k''_x f_5 + k'_x f_4}{k_0 |\varepsilon_m|^2} - (k'_{z,m} + k_0 k_m) f_7 + (k''_{z,m} + k_0 n_m) f_6, \tag{S8}$$

$$f_3(x,z,\delta) = k'_x x + (k''_{z,m} + k_0 n_m) z + \delta, \tag{S9}$$

$$f_4 = \varepsilon'_m \left[ k'_x (n_m + n_m^2 + k_m^2) + k''_x k_m \right] + \varepsilon''_m \left[ -k'_x k_m + k''_x (n_m + n_m^2 + k_m^2) \right], \tag{S10}$$

$$f_5 = \varepsilon'_m \left[ k'_x k_m - k''_x (n_m + n_m^2 + k_m^2) \right] + \varepsilon''_m \left[ k'_x (n_m + n_m^2 + k_m^2) + k''_x k_m \right], \tag{S11}$$

$$f_6 = 1 + n_m + \frac{\varepsilon'_m \left[ k'_{z,m} k_m - k''_{z,m} (n_m + n_m^2 + k_m^2) \right]}{k_0 |\varepsilon_m|^2} + \frac{\varepsilon''_m \left[ k'_{z,m} (n_m + n_m^2 + k_m^2) + k''_{z,m} k_m \right]}{k_0 |\varepsilon_m|^2}, \tag{S12}$$

$$f_7 = -k_m - \frac{\varepsilon'_m \left[ k'_{z,m} (n_m + n_m^2 + k_m^2) + k''_{z,m} k_m \right]}{k_0 |\varepsilon_m|^2} + \frac{\varepsilon''_m \left[ k'_{z,m} k_m - k''_{z,m} (n_m + n_m^2 + k_m^2) \right]}{k_0 |\varepsilon_m|^2}, \tag{S13}$$

where $\varepsilon_m = \varepsilon'_m + i\varepsilon''_m = \tilde{n}_m^2$ is the dielectric function of a metal. The term $Q_{las\text{-}SPP}$ is responsible for the periodic laser energy absorption along the lateral distance $x$ and, therefore, for LIPSS formation.

The third term of the Eq. (S1) is a result of the SPP-SPP interference:

$$Q_{SPP\text{-}SPP}(\mathbf{r},t) = \beta^2 F_{inc} f_8 \Phi_2(y) G_2(x,t) \exp\left(2(k'_{z,m} z - k''_x x)\right), \tag{S14}$$

where

$$f_8 = \frac{2\left[\varepsilon'_m (k'_x k''_x - k'_{z,m} k''_{z,m}) + \varepsilon''_m (k''^2_x + k'^2_{z,m})\right]}{k_0 |\varepsilon_m|^2}. \tag{S15}$$

$$\Phi_2(y) = \theta\left(-y + \frac{a_{\text{beam}}}{2}\right)\theta\left(y + \frac{a_{\text{beam}}}{2}\right), \tag{S16}$$

$$G_2(x,t) = \frac{1}{\tau}\sqrt{\frac{\sigma}{\pi}} e^{-\sigma \frac{\left(x - v_{g,\text{SPP}}(t-t_0)\right)^2}{v_{g,\text{SPP}}^2 \tau^2}}. \tag{S17}$$

The term $Q_{\text{SPP-SPP}}$ describes the propagation and decay of SPP after the action of the laser pulse.

**References**

1. P.N. Terekhin, O. Benhayoun, S.T. Weber, D.S. Ivanov, M.E. Garcia, B. Rethfeld, Influence of surface plasmon polaritons on laser energy absorption and structuring of surfaces, Appl. Sur. Sci. **512**, 144420 (2020).

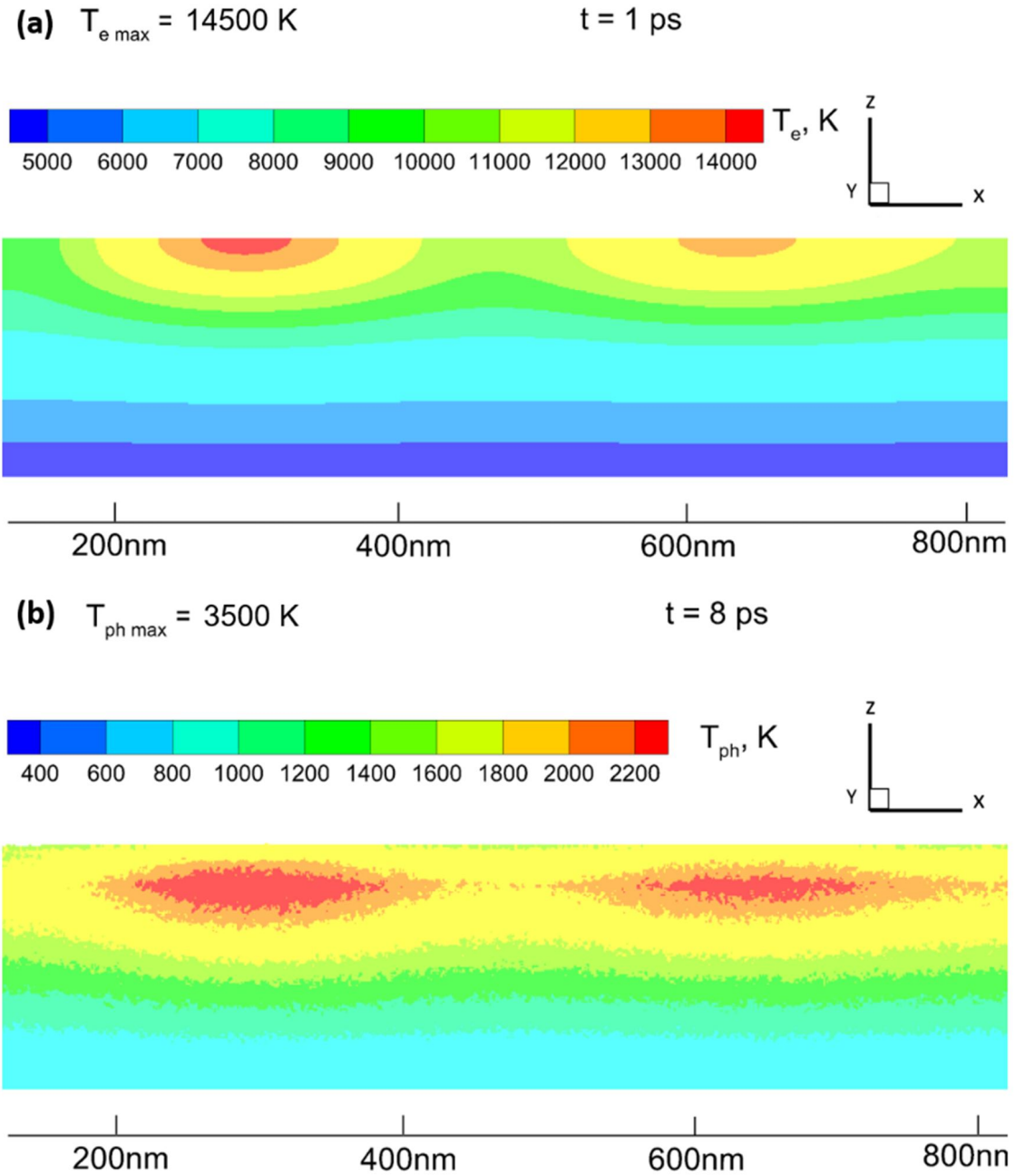

**FIG. S1.** (a) Electron and (b) phonon temperature fields are shown at 1 ps and 8 ps, respectively. The maximum temperature values developed during the simulation of the laser pulse interacting with Au at the incident fluence of 130 mJ/cm$^2$ are indicated.

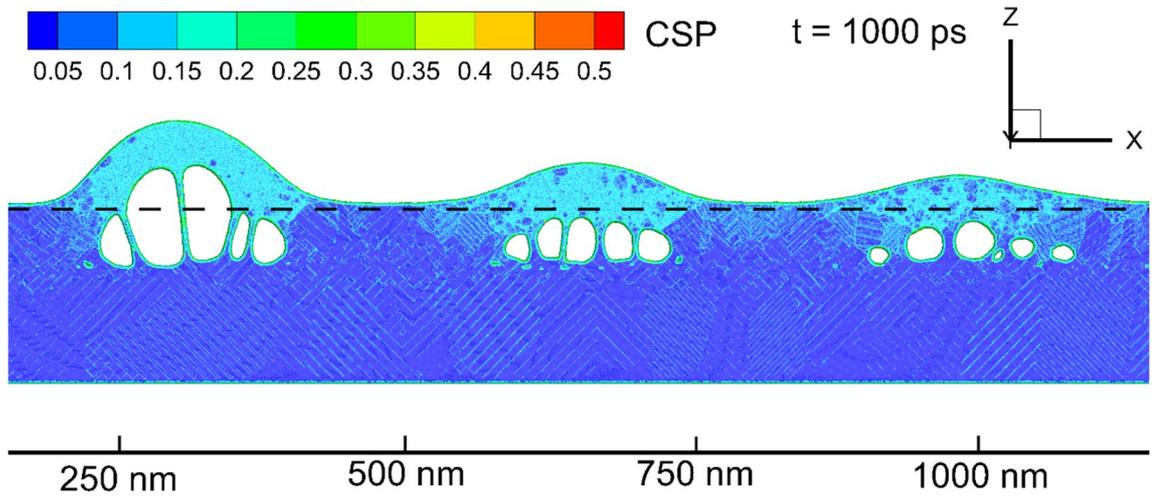

**FIG. S2.** Zoomed atomic snapshot of the Au target at 1000 ps after the pulse for the case of an incident fluence of 130 mJ/cm$^2$. The atoms are colored by the Central Symmetry Parameter (CSP) for distinguishing the local crystal structure as follow: crystal < 0.08 < defects (dislocations) < 0.11 < liquid < 0.25 < surfaces < 0.50 < vapor (free atoms). The dashed line indicates the sample surface before the irradiation.

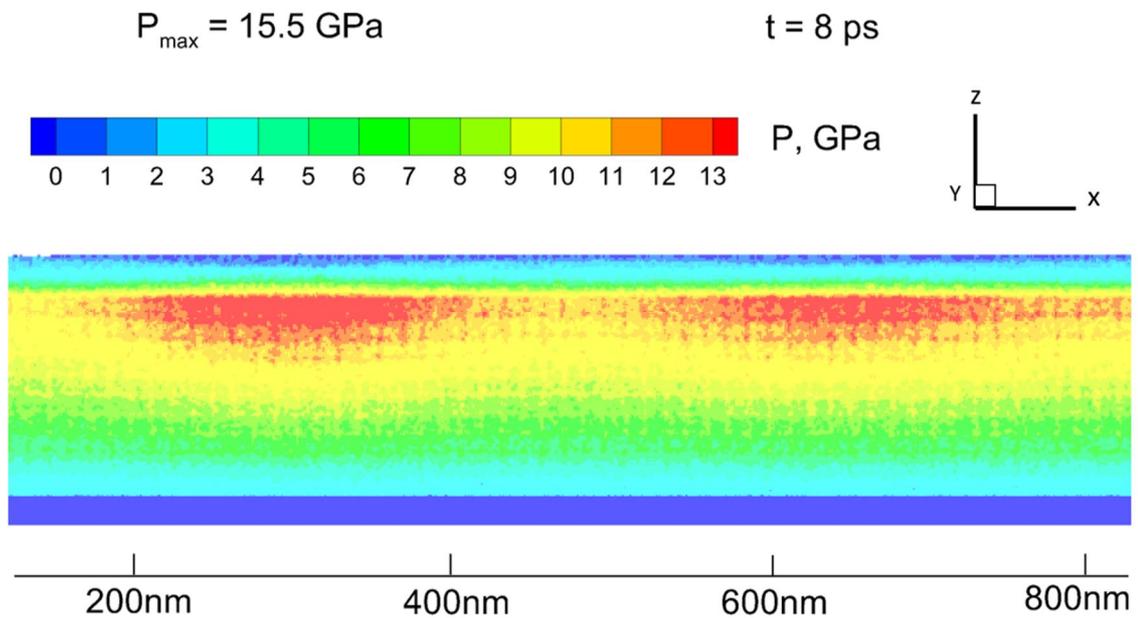

**FIG. S3.** Pressure distribution is shown at 8 ps. The maximum pressure value is indicated.